\documentclass{article}

\usepackage[preprint, nonatbib]{neurips_2019}

\usepackage[utf8]{inputenc} 
\usepackage[T1]{fontenc}    
\usepackage{hyperref}       
\usepackage{url}            
\usepackage{booktabs}       
\usepackage{amsfonts}       
\usepackage{nicefrac}       
\usepackage{microtype}      
\usepackage{caption}
\usepackage{subcaption}
\usepackage{graphicx}
\usepackage{textcomp}
\usepackage{caption} 

\usepackage[utf8]{inputenc} 
\usepackage{hyperref}       
\usepackage{url}            
\usepackage{booktabs}       
\usepackage{amsfonts}       
\usepackage{amsmath}
\usepackage{graphicx,wrapfig,lipsum}
\usepackage{amsfonts}
\usepackage{booktabs}
\usepackage{empheq}
\usepackage{amsthm}
\usepackage{caption}
\usepackage{subcaption}
\usepackage{mathrsfs}  
\usepackage{amssymb}
\usepackage{tikz}
\usetikzlibrary{calc,positioning}

\theoremstyle{plain}

\theoremstyle{definition}

\newcommand{\nomath}[1]{\ifmmode
\mathrm{#1}%
\else
#1%
\fi}

\newcommand{\TODO}[1][]{\textcolor{red}{\ifx&#1&
\nomath{TODO}
\else
\nomath{TODO: #1}
\fi}}

\title{Deep Learning for Estimating Synaptic Health of Primary Neuronal Cell Culture}

\author{ 
	Andrey Kormilitzin$^{1, 2, }$\thanks{Equal contribution.}
	\And
	Xinyu Yang$^{1, 2, }$\footnotemark[1]
	\And
	William H. Stone$^{1, }$\footnotemark[1]
	\And
	Caroline Woffindale$^{2, }$\footnotemark[1]
	\And
	Francesca Nicholls$^{2 }$
	\And
	Elena Ribe$^{2 }$
	\And
	Alejo Nevado-Holgado$^{2 }$
	\And
	Noel Buckley$^{2 }$
	\And
	\AND \\[-12pt]
	\null$^1$ Mathematical Institute, University of Oxford \\
	\null$^2$ Department of Psychiatry, University of Oxford \\
	\texttt{william.stone2@queens.ox.ac.uk}\\
    \texttt{$\{$andrey.kormilitzin, xinyu.yang, caroline.woffindale, francesca.nicholls,}\\
    \texttt{elena.ribe, alejo.nevado-holgado, noel.buckley$\}$@\hspace{0.1pt}psych.ox.ac.uk}
}

\begin{document}

\maketitle

\begin{abstract}
  Understanding the morphological changes of primary neuronal cells induced by chemical compounds is essential for drug discovery. Using the data from a single high-throughput imaging assay, a classification model for predicting the biological activity of candidate compounds was introduced. The image recognition model which is based on deep convolutional neural network (CNN) architecture with residual connections achieved accuracy of 99.6$\%$ on a binary classification task of distinguishing untreated and treated rodent primary neuronal cells with Amyloid-$\beta_{(25-35)}$. 
  

\end{abstract}

\section{Introduction}

Alzheimer's disease is the third leading cause of death in the US and globally \cite{james2014contribution}. The economic cost of Alzheimer's disease is estimated to be $\$$36.512 billion in the US alone and expected to rise to $\$$2 trillion by 2030 globally \cite{luengo2015uk}. It has been shown that the $\beta$-Amyloid (A$\beta$) peptide is involved in Alzheimer's disease \cite{mckhann2011diagnosis} and found in the brains of Alzheimer patients \cite{hamley2012amyloid, westwood2018discovery}. Searching for new chemical compounds, which moderate the previously identified A$\beta$ driven DKK1-Wnt-PCP-JNK neurotoxic signalling pathway, may reverse the effect of toxic depositions of A$\beta$ plaques in the brain and will inform the development of novel therapeutic approaches for the treatment of Alzheimer's disease. 
We hypothesise that by reversing or preventing these changes in gene expression, this may protect against the toxic effects seen in primary rodent neuronal cultures following treatment with A$\beta_{(25-35)}$. The standard and laborious approach to assess whether a candidate compound may have a protective effect, is to perform a statistical hypothesis testing using the features extracted with CellProflier image processing tool across various treatment regimes. The proposed deep-learning approach eliminates the need for this statistical comparison and estimates the treatment effects from raw pixels of high-throughput images.

\section{Related work}

Recent years have seen an explosion of applications of the deep learning methods to medical imaging, including computer-aided diagnosis (CAD) in radiology and medical image analysis \cite{suzuki2017overview, liu2017detecting, liuartificial}. The efficiency of deep learning models for cytometry \cite{gupta2019deep} has been widely recognised and applied to cell imaging \cite{komura2018machine}, virtual staining with generative adversarial networks (GAN) \cite{rivenson2018deep, rivenson2019virtual}, fluorescence microscopy \cite{wang2019deep} and reconstructing cell cycles and disease progression \cite{eulenberg2017reconstructing}. However, despite the wide popularity and maturity of the deep learning approach, very little has been done to estimate the effect of biological activity of neuronal cells induced by compounds and searching for drugs that may protect against neurodegeneration and Alzheimer's disease. Simm \textit{et al.} in \cite{simm2018repurposing} suggested to re-purpose high-throughput images assay to predict biological activity in drug discovery, however this approach depends on the features extracted from CellProfiler \cite{carpenter2006cellprofiler} and lacks the flexibility of the CNN models \cite{lecun2015deep} which learn features directly from raw pixels of images.



\section{Materials and methods}

\subsection{The assay design}

Primary neuronal cells were seeded in 96 well plates and cultured for 21 days \textit{in vitro} to allow them to establish neuronal connections and to synaptically mature. Cells are then pre-treated for one hour with one candidate compound in doses of either 10, 3 or 1 $\mu$M\footnote{$\mu$M stands for micro molar, is a measure of the concentration of a chemical substance in a solution} before the addition of 30 $\mu$M of A$\beta_{(25-35)}$ or a vehicle control. Cells are incubated for a further 48 hours before they are fixed and stained for various markers. The assay plate configuration and the location of compounds and A$\beta$ at various doses (in $\mu$M) are presented in Fig. \ref{fig:assay_plate} and Tab. \ref{tab:compounds} respectively. For example, wells in column 'B' at rows 2, 3, 4 and wells in columns 'C' at rows 5,6,7 all contain zero dose of a compound and a zero dose of $A\beta$, which corresponds to a vehicle control. Wells in column 'D' and rows 2,3,4 and wells in column 'E' and rows 5,6,7 all contain a combination of 1 $\mu M$ of a compound and 30 $\mu M$ of $A\beta$.

\begin{minipage}[b]{0.45\textwidth}
    \centering
        \includegraphics[scale=0.37]{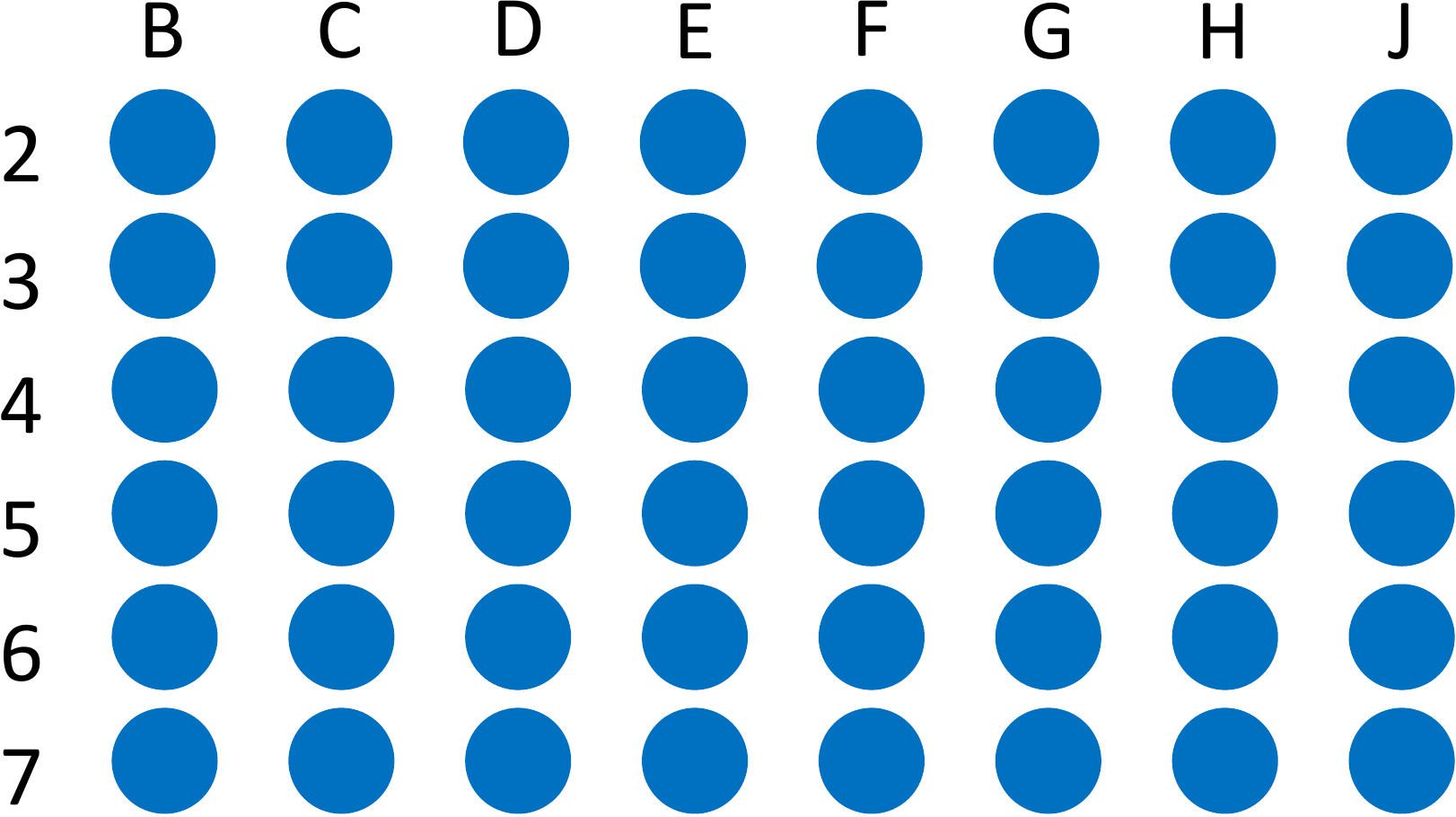}
    \captionof{figure}{An assay plate with wells.}
    \label{fig:assay_plate}
\end{minipage}
  \hfill
\begin{minipage}[b]{0.55\textwidth}
    \centering
        \begin{tabular}{c|r|r}
          Well position index & Compound & $A\beta$\\ \hline
            B\{2,3,4\}, C\{5,6,7\} & 0 & 0\\
            C\{2,3,4\}, D\{5,6,7\} & 0 & 30\\
            D\{2,3,4\}, E\{5,6,7\} & 1 & 0\\
            E\{2,3,4\}, F\{5,6,7\} & 1 & 30\\
            F\{2,3,4\}, G\{5,6,7\} & 3 & 0\\
            G\{2,3,4\}, H\{5,6,7\} & 3 & 30\\
            H\{2,3,4\}, J\{5,6,7\} & 10 & 0\\
            F\{2,3,4\}, B\{5,6,7\} & 10 & 30\\\hline
        \end{tabular}
    \captionof{table}{The position map of compounds and doses.}
    \label{tab:compounds}
\end{minipage}

To highlight specific cellular structure, four stains were used: MAP2 (neuronal cytoskeleton marker), PSD-95 (post-synaptic protein marker), Synaptophysin (pre-synaptic protein marker) and Hoescht 33342 (nucleic marker). Cells are then imaged at magnification of 40x using the high-content GE IN Cell Analyzer 6000\textsuperscript{TM} imaging system.  Each well was images 30 times ('field views') using non-overlapping strides for each of the four channels; MAP2 (Cy5), Nucleus (DAPI), Synaptophysin (dsRed), PSD-95 (FITC). This resulted in 5760 images per plate used in the analysis.


\subsection{The deep learning model}

The objective of this work was to develop a classification model able to reliably recognise morphological changes in neuronal cells treated with A$\beta_{(25-35)}$ versus the vehicle control ('untreated'). Once trained to a satisfactory accuracy, the the model was applied to screen candidate compounds and determine whether images of cells treated with a compound + A$\beta_{(25-35)}$ were more similar to those treated with A$\beta_{(25-35)}$ alone. For this purpose, we developed a CNN-based \cite{lecun2015deep} binary classifier. The model used in classification is based on the Resnet18 \cite{he2016deep} architecture with weights pre-trained on the ImageNet data \cite{imagenet_cvpr09} and implemented in PyTorch \cite{paszke2017automatic} and fastai \cite{howard2018fastai} python libraries. The model was fine-tuned on a binary classification task using 6480 of untreated and 6480 of treated images of neuronal cells of size 2048 by 2048 pixels. To mitigate the problem of overfitting, data augmentations with default settings of random cropping, flipping both horizontally and vertically, small rotations and shearing were used. We used dropout \cite{hinton2012improving} of 0.35, leading to stable convergence. The training was done using SGD with momentum of 0.9, slanted learning rate \cite{smith2018disciplined, howard2018universal} with the optimal one \cite{li2018visualizing}. The batch size of 4 was used due to memory limitation of NVIDIA 2080Ti GPU. The entire training procedure was split into two bits: first training the classification head and then gradual unfreezing of convolutional layers. The classification head comprised one fully connected layers of 512 neurons with ReLU activation function \cite{ramachandran2017searching} and the final binary softmax function. No weight decay was used.

In order to visualise individual pixels that supported the decision of the classification model and for sanity check that the model's decisions are not due to image artefacts the weighted Grad-CAM \cite{selvaraju2017grad} was implemented. The visualisation methods showed that the “attention” of the model was drawn towards specific neurites; providing assurance that the model fit true features of the data.

A$\beta$ peptide has been found as a dominant mediator of neuronal dystrophy and synaptic loss \cite{henriques2015abeta} which is reflected in the morphology of neuronal cytoskeleton structure. Therefore in the current work we focused only the Cy5 stain which highlights the cytoskeleton structure. However, we are planning on extending this approach and develop a multi-channel model which learns from all four stains simultaneously. We believe that further inclusion of all four channels (Cy5, DAPI, dsRed and FITC) will improve the detection capability and improve our understanding of morphological changes induced by compounds at various doses.

\section{Results}

\begin{figure}[ht]
    \centering
    \includegraphics[width=0.99\textwidth]{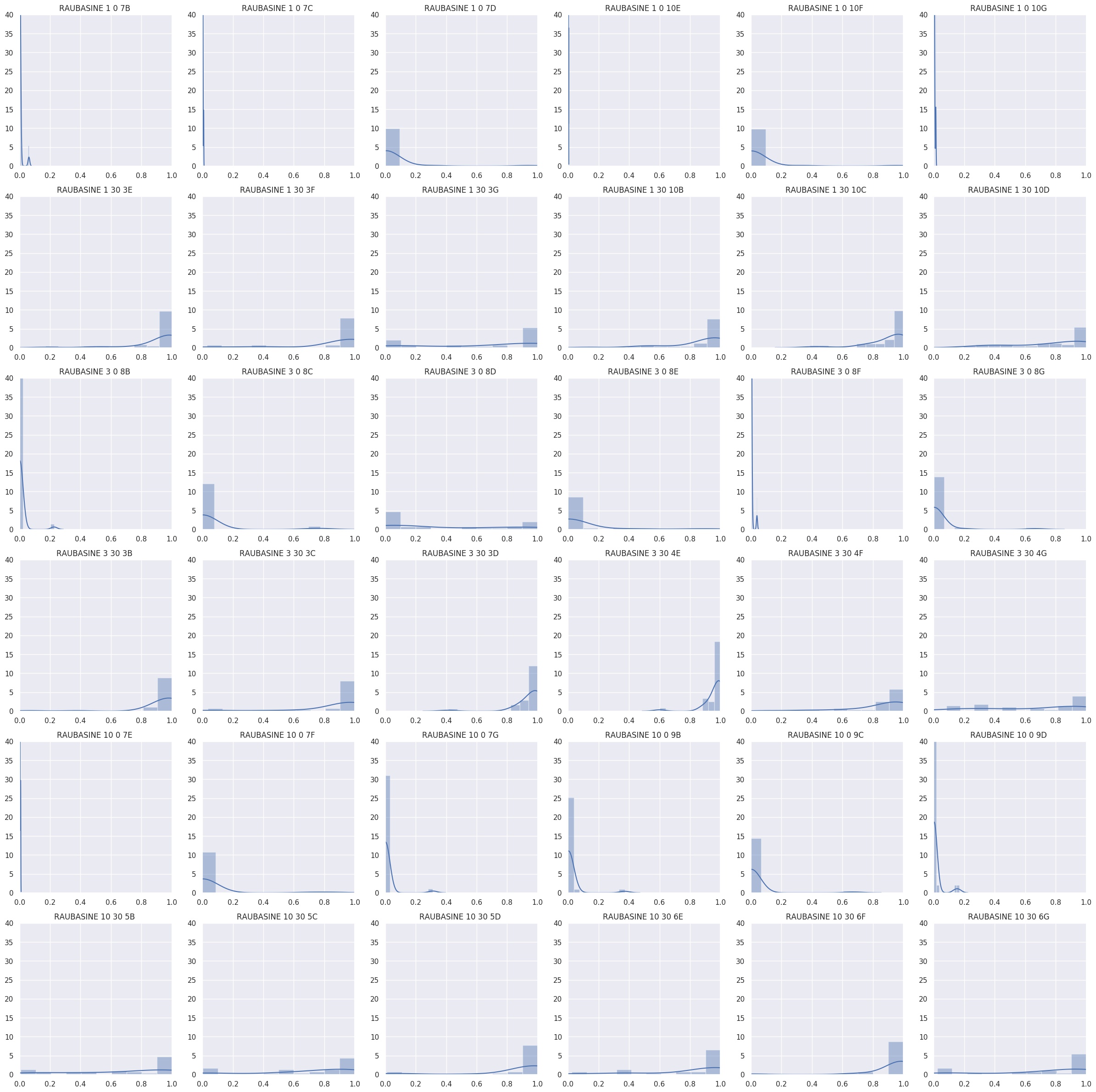}
    \caption{A distribution of the predicted scores for \textbf{Raubasine}.}
    \label{fig:distribution_of_predictions_raubasine}
\end{figure}

\begin{figure}[ht]
    \centering
    \includegraphics[width=0.99\textwidth]{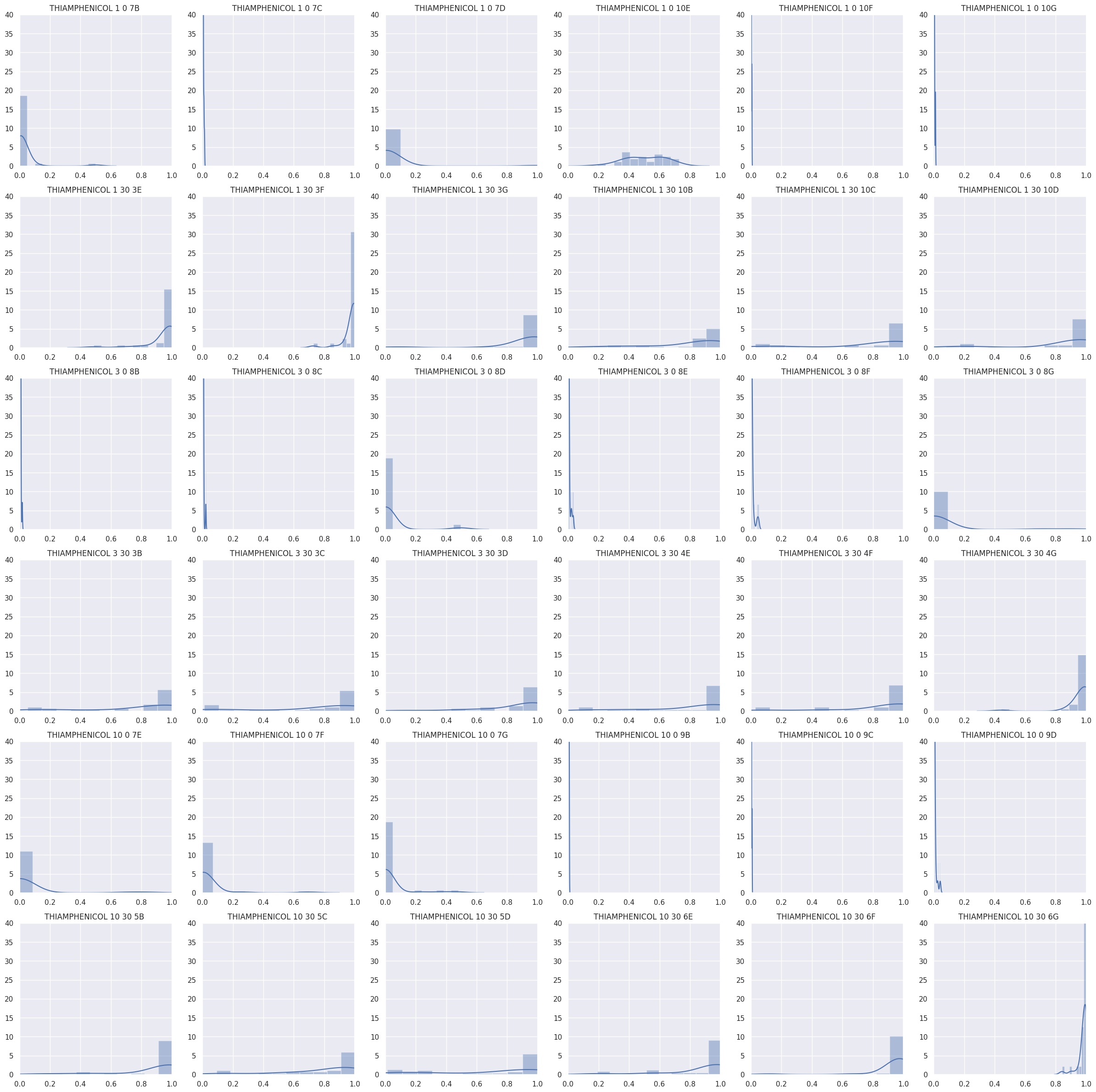}
    \caption{A distribution of the predicted scores for \textbf{Thiamphenicol}. Each column corresponds to a single compound configuration. Six columns correspond to.}
    \label{fig:distribution_of_predictions_thiamphenicol}
\end{figure}

Of 36 assay plates, with a unique compound in each plate, 34 were randomly selected for training and the remaining 2 for testing. However, the model was trained only on images of neurons treated with vehicle controls and $A\beta$, whereas all remaining wells with various treatment regimes were unused. Therefore the unused images can be safely used for inference. The main reason behind setting aside two assay plates was to test the model performance on vehicle controls and $A\beta$. The list of 36 compounds is presented in Table \ref{tab:list_of_compounds} in \ref{appendix:c}. and the distribution of predicted scored of vehicle controls and $A\beta$ are presented in Fig. \ref{fig:distribution_of_predictions_raubasine} and Fig. \ref{fig:distribution_of_predictions_thiamphenicol}.

We applied data augmentation to training images including random scaling, flipping, cropping, and other affine transformations using the package imgaug \cite{imgaug}. The model was trained for 10 epochs on the last convolution block and the fully connected layer with learning rate \texttt{1e-3}, and another 10 epochs on all layers with learning rate \texttt{1e-4}, all using the stochastic gradient descent (SGD) optimizer with momentum 0.9. The training results are summarised in Table \ref{tab:training_results}. 

\begin{table}[ht]
    \centering
    \begin{tabular}{c|c|c|c}
        Training loss & Training accuracy & Validation loss & Validation accuracy\\\hline
        0.012 & 99.68$\%$ & 0.013 & 99.58$\%$\\\hline
    \end{tabular}
    \caption{Training and validation accuracies and losses.}
    \label{tab:training_results}
\end{table}

In the inference mode, the model prediction scores for each individual well were averaged over field views. The resulting prediction for each treatment regimen was averaged over six wells and the standard deviation computed. The results of model predictions for Raubasine and Thiamphenicol compounds are presented in Tables \ref{tab:raubasine} and \ref{tab:thiamphenicol} respectively. The threshold of 0.5 was used to convert predicted scores to dichotomous categories (untreated and treated with 30 $
\mu M$ of $A\beta_{(25-35)}$).

\begin{table}[htpb]
\centering
    \begin{tabular}{l|l|l|l|l|l|l|l|l}
        compound dose ($\mu$M)                  &	0   &	0   &   1   &	1   &   3   &	3   &   10   &	10  \\
        A$\beta$ dose ($\mu$M)                  &	0   &	30  &	0   &	30  &	0   &	30	&	0    &  30  \\\hline\hline
        well$_1$                                & 0.0   &   0.87&   0.01& 0.92  &   0.01&  0.93 &   0.01 & 0.69 \\  	
        well$_2$                                & 0.01  &   0.93&   0.0 & 0.83  &   0.07&  0.86 &   0.07 & 0.68 \\  
        well$_3$                                & 0.21  &   0.87&   0.04& 0.7   &   0.47&  0.94 &   0.01 & 0.87 \\
        well$_4$                                & 0.08  &   0.91&   0.0 & 0.82  &   0.07&  0.96 &   0.02 & 0.81 \\
        well$_5$                                & 0.03  &   0.93&   0.06& 0.87  &   0.01&  0.84 &   0.03 & 0.91 \\
        well$_6$                                & 0.0   &   0.89&   0.0 & 0.48  &   0.02&  0.68 &   0.04 & 0.71 \\\hline
        mean predicted score   & \textbf{0.05}  &   \textbf{0.9} &   \textbf{0.02}& \textbf{0.77}  &   \textbf{0.10}&  \textbf{0.87} &   \textbf{0.03} & \textbf{0.78} \\
        standard deviation                      & 0.11  &   0.21&   0.07& 0.27  &   0.14&  0.18 &   0.09 & 0.29 \\ \hline
    \end{tabular}
    \caption{An example of screening of Raubasine compound at various doses.}
    \label{tab:raubasine}
\end{table}

The combination of zero dose of both a compound and $A\beta$ corresponds to a vehicle control. For example, the neuronal cells treated with Raubasine 1 $\mu M$ and 30 $\mu M$ of $A\beta$, were classified as treated with $A\beta$ 30 $\mu M$, which means there is a not significant protective effect of Raubasine at this dose against synaptic loss induced by $A\beta$. Similarly, neuronal cells treated with larger doses of Raubasine (10 $\mu M$) were also classified as treated with only $A\beta$. 

\begin{table}[htpb]
\centering
    \begin{tabular}{l|l|l|l|l|l|l|l|l}
        compound dose ($\mu$M)                  &	0   &	0   &   1       &	1   &   3   &	3   &   10   &	10  \\
        A$\beta$ dose ($\mu$M)                  &	0   &	30  &	0       &	30  &	0   &	30	&	0    &  30  \\\hline\hline
        well$_1$                                & 0.01  &   0.83&   0.02    & 0.95  &   0.01&  0.78 &   0.05 & 0.87 \\  	
        well$_2$                                & 0.02  &   0.83&   0.01    & 0.96  &   0.01&  0.76 &   0.03 & 0.79 \\  
        well$_3$                                & 0.03  &   0.93&   0.08    & 0.9   &   0.04&  0.86 &   0.02 & 0.68 \\
        well$_4$                                & 0.0   &   0.84&   0.18    & 0.76  &   0.01&  0.81 &   0.01 & 0.85 \\
        well$_5$                                & 0.0   &   0.86&   0.0     & 0.81  &   0.01&  0.85 &   0.01 & 0.94 \\
        well$_6$                                & 0.0   &   0.88&   0.01    & 0.85  &   0.06&  0.97 &   0.02 & 0.98 \\\hline
        mean predicted score    &\textbf{0.01}&\textbf{0.86}&\textbf{0.05}&\textbf{0.87}&\textbf{0.02}&\textbf{0.84}&\textbf{0.02}&\textbf{0.85} \\
        standard deviation                      & 0.03  &   0.22&   0.07    & 0.19  &   0.07&  0.23 &   0.06 & 0.2 \\ \hline
    \end{tabular}
    \caption{An example of screening of Thiamphenicol compound at various doses.}
    \label{tab:thiamphenicol}
\end{table}

\section{Conclusions}

We report the validation accuracy of 99.58$\%$ on a binary images classification problem using high-throughput images of neuronal cells that have been treated with vehicle control versus images treated with 30 $\mu M$ of A$\beta_{(25-35)}$. Amyloid-$\beta$ is thought to induce synapse loss. The developed deep learning based high throughput method which is able to identify and measure this synapse loss and then use this technique to identify when compounds can rescue it. 

None of 36 screened compounds (Table \ref{tab:list_of_compounds}) applied to neuronal cells at various doses was found as having a substantial protective effect against $A\beta_(25-35)$. This result was confirmed by screening the same assay plates of 36 compounds using the image processing pipeline of CellProfiler followed by statistical hypothesis testing of the extracted features.

In contrast to a deep CNN model, CellProfiler \cite{carpenter2006cellprofiler} requires a substantial amount of computational resource to process the same number of images. For example, with 48 CPUs, a typical imaging processing of a single plate takes approximately 15 hours to complete in contrast to 60 seconds using the deep learning model in the inference mode. In addition to a long computation time, large scale screening of hundreds or thousands of compounds will result in a substantial carbon footprint compared to re-suing of pre-trained deep learning models.

The visualisation of pixel-wise attention of the model using Grad-CAM is presented in \ref{appendix:a}. The results demonstrated in Fig. \ref{fig:raubasine_grad_cam} and Fig. \ref{fig:thiamphenicol_grad_cam} are randomly selected images from each of the treatment regimes for two compounds (Raubasine and Thiamphenicol) which were not used in the training (hold-out set). Various bits of the cytoskeleton structure found important for the classification decision of the model. The Grad-CAM methods highlights the important regions in the image for predicting the class label.

The predicted scores are predominantly distributed around zero or one, except for a few wells (\ref{appendix:b}). This chart allows neurobiologists to manually inspect particular wells where the distribution of predicted scores is not clustered around zero or one. Such cases might be caused by contaminated wells, blurred imaging or other artefacts which may confuse the image classification model.

Learning synaptic features and segmentation of cytoskeleton structure have been explored previously \cite{berning2015segem, plaza2018analyzing, matejeksynapse}. Recently, a signature-based approach \cite{lyons2014rough, chevyrev2016primer} to learning from ordered data has been proposed and applied to various tasks in image and character recognition \cite{graham2013sparse, xie2017learning, yang2017leveraging}. The signature transforms sequentially ordered data - a path, into a succinct representation, effectively capturing global and local information. The signature transformation is used as feature representation in downstream machine learning tasks. The signature method was used in mental health \cite{kormilitzin2016application, kormilitzin2017detecting, arribas2017signature}, prediction of diagnosis of Alzheimer's disease \cite{moore2018using, moore2019random}, finance \cite{arribas2018derivatives, gyurko2013extracting, levin2013learning} and deep signature learning \cite{bonnier2019deep}. The cytoskeleton structure could be factorised into a collection of branching diagrams, making it suitable for signature-based learning. We plan to investigate this approach to effective learning of morphological changes of neuronal cells in our future works.

Although the images of neuronal cells were taken from rodents, potentially the developed method could be generalised in humans via transfer learning with the final goal of efficiently identifying novel compounds to reverse the effect of A$\beta$ and drug development for neurodegeneration and Alzheimer's disease.

\section*{Acknowledgements}
AK is supported by the MRC under the Pathfinder programme grant MC/PC/17215. The work of XY is supported in part by the Oxford Alzheimer's Researck UK (ARUK) Pilot Project Award.


\bibliographystyle{unsrt}

\nocite{*}
\begin{thebibliography}{10}

\bibitem{james2014contribution}
Bryan~D James, Sue~E Leurgans, Liesi~E Hebert, Paul~A Scherr, Kristine Yaffe,
  and David~A Bennett.
\newblock Contribution of alzheimer disease to mortality in the united states.
\newblock {\em Neurology}, 82(12):1045--1050, 2014.

\bibitem{luengo2015uk}
Ramon Luengo-Fernandez, Jose Leal, and Alastair Gray.
\newblock Uk research spend in 2008 and 2012: comparing stroke, cancer,
  coronary heart disease and dementia.
\newblock {\em BMJ open}, 5(4):e006648, 2015.

\bibitem{mckhann2011diagnosis}
Guy~M McKhann, David~S Knopman, Howard Chertkow, Bradley~T Hyman, Clifford~R
  Jack~Jr, Claudia~H Kawas, William~E Klunk, Walter~J Koroshetz, Jennifer~J
  Manly, Richard Mayeux, et~al.
\newblock The diagnosis of dementia due to alzheimer’s disease:
  Recommendations from the national institute on aging-alzheimer’s
  association workgroups on diagnostic guidelines for alzheimer's disease.
\newblock {\em Alzheimer's \& dementia}, 7(3):263--269, 2011.

\bibitem{hamley2012amyloid}
Ian~W Hamley.
\newblock The amyloid beta peptide: a chemist’s perspective. role in
  alzheimer’s and fibrillization.
\newblock {\em Chemical reviews}, 112(10):5147--5192, 2012.

\bibitem{westwood2018discovery}
Sarah Westwood, Alison~L Baird, Sneha~N Anand, Liu Shi, Alejo~J Nevado-Holgado,
  Andrey Kormilitzin, Abdul Hye, Nicholas~J Ashton, Angharad Morgan, Samuel
  Touchard, et~al.
\newblock Discovery, replication and extension study of plasma proteomic
  biomarkers relating to brain amyloid burden and alzheimer’s disease
  progression.
\newblock {\em Alzheimer's \& Dementia: The Journal of the Alzheimer's
  Association}, 14(7):P201--P202, 2018.

\bibitem{suzuki2017overview}
Kenji Suzuki.
\newblock Overview of deep learning in medical imaging.
\newblock {\em Radiological physics and technology}, 10(3):257--273, 2017.

\bibitem{liu2017detecting}
Yun Liu, Krishna Gadepalli, Mohammad Norouzi, George~E Dahl, Timo Kohlberger,
  Aleksey Boyko, Subhashini Venugopalan, Aleksei Timofeev, Philip~Q Nelson,
  Greg~S Corrado, et~al.
\newblock Detecting cancer metastases on gigapixel pathology images.
\newblock {\em arXiv preprint arXiv:1703.02442}, 2017.

\bibitem{liuartificial}
Yun Liu, Timo Kohlberger, Mohammad Norouzi, George~E Dahl, Jenny~L Smith, Arash
  Mohtashamian, Niels Olson, Lily~H Peng, Jason~D Hipp, and Martin~C Stumpe.
\newblock Artificial intelligence--based breast cancer nodal metastasis
  detection: Insights into the black box for pathologists.
\newblock {\em Archives of Pathology \& Laboratory Medicine}.

\bibitem{gupta2019deep}
Anindya Gupta, Philip~J Harrison, H{\aa}kan Wieslander, Nicolas Pielawski,
  Kimmo Kartasalo, Gabriele Partel, Leslie Solorzano, Amit Suveer, Anna~H
  Klemm, Ola Spjuth, et~al.
\newblock Deep learning in image cytometry: a review.
\newblock {\em Cytometry Part A}, 95(4):366--380, 2019.

\bibitem{komura2018machine}
Daisuke Komura and Shumpei Ishikawa.
\newblock Machine learning methods for histopathological image analysis.
\newblock {\em Computational and structural biotechnology journal}, 16:34--42,
  2018.

\bibitem{rivenson2018deep}
Yair Rivenson, Hongda Wang, Zhensong Wei, Yibo Zhang, Harun Gunaydin, and
  Aydogan Ozcan.
\newblock Deep learning-based virtual histology staining using
  auto-fluorescence of label-free tissue.
\newblock {\em arXiv preprint arXiv:1803.11293}, 2018.

\bibitem{rivenson2019virtual}
Yair Rivenson, Hongda Wang, Zhensong Wei, Kevin de~Haan, Yibo Zhang, Yichen Wu,
  Harun G{\"u}nayd{\i}n, Jonathan~E Zuckerman, Thomas Chong, Anthony~E Sisk,
  et~al.
\newblock Virtual histological staining of unlabelled tissue-autofluorescence
  images via deep learning.
\newblock {\em Nature biomedical engineering}, 3(6):466, 2019.

\bibitem{wang2019deep}
Hongda Wang, Yair Rivenson, Yiyin Jin, Zhensong Wei, Ronald Gao, Harun
  G{\"u}nayd{\i}n, Laurent~A Bentolila, Comert Kural, and Aydogan Ozcan.
\newblock Deep learning enables cross-modality super-resolution in fluorescence
  microscopy.
\newblock {\em Nat. Methods}, 16:103--110, 2019.

\bibitem{eulenberg2017reconstructing}
Philipp Eulenberg, Niklas K{\"o}hler, Thomas Blasi, Andrew Filby, Anne~E
  Carpenter, Paul Rees, Fabian~J Theis, and F~Alexander Wolf.
\newblock Reconstructing cell cycle and disease progression using deep
  learning.
\newblock {\em Nature communications}, 8(1):463, 2017.

\bibitem{simm2018repurposing}
Jaak Simm, G{\"u}nter Klambauer, Adam Arany, Marvin Steijaert, J{\"o}rg~Kurt
  Wegner, Emmanuel Gustin, Vladimir Chupakhin, Yolanda~T Chong, Jorge Vialard,
  Peter Buijnsters, et~al.
\newblock Repurposing high-throughput image assays enables biological activity
  prediction for drug discovery.
\newblock {\em Cell chemical biology}, 25(5):611--618, 2018.

\bibitem{carpenter2006cellprofiler}
Anne~E Carpenter, Thouis~R Jones, Michael~R Lamprecht, Colin Clarke, In~Han
  Kang, Ola Friman, David~A Guertin, Joo~Han Chang, Robert~A Lindquist, Jason
  Moffat, et~al.
\newblock Cellprofiler: image analysis software for identifying and quantifying
  cell phenotypes.
\newblock {\em Genome biology}, 7(10):R100, 2006.

\bibitem{lecun2015deep}
Yann LeCun, Yoshua Bengio, and Geoffrey Hinton.
\newblock Deep learning.
\newblock {\em nature}, 521(7553):436, 2015.

\bibitem{he2016deep}
Kaiming He, Xiangyu Zhang, Shaoqing Ren, and Jian Sun.
\newblock Deep residual learning for image recognition.
\newblock In {\em Proceedings of the IEEE conference on computer vision and
  pattern recognition}, pages 770--778, 2016.

\bibitem{imagenet_cvpr09}
J.~Deng, W.~Dong, R.~Socher, L.-J. Li, K.~Li, and L.~Fei-Fei.
\newblock {ImageNet: A Large-Scale Hierarchical Image Database}.
\newblock In {\em CVPR09}, 2009.

\bibitem{paszke2017automatic}
Adam Paszke, Sam Gross, Soumith Chintala, Gregory Chanan, Edward Yang, Zachary
  DeVito, Zeming Lin, Alban Desmaison, Luca Antiga, and Adam Lerer.
\newblock Automatic differentiation in pytorch.
\newblock 2017.

\bibitem{howard2018fastai}
Jeremy Howard et~al.
\newblock fastai.
\newblock \url{https://github.com/fastai/fastai}, 2018.

\bibitem{hinton2012improving}
Geoffrey~E Hinton, Nitish Srivastava, Alex Krizhevsky, Ilya Sutskever, and
  Ruslan~R Salakhutdinov.
\newblock Improving neural networks by preventing co-adaptation of feature
  detectors.
\newblock {\em arXiv preprint arXiv:1207.0580}, 2012.

\bibitem{smith2018disciplined}
Leslie~N Smith.
\newblock A disciplined approach to neural network hyper-parameters: Part
  1--learning rate, batch size, momentum, and weight decay.
\newblock {\em arXiv preprint arXiv:1803.09820}, 2018.

\bibitem{howard2018universal}
Jeremy Howard and Sebastian Ruder.
\newblock Universal language model fine-tuning for text classification.
\newblock {\em arXiv preprint arXiv:1801.06146}, 2018.

\bibitem{li2018visualizing}
Hao Li, Zheng Xu, Gavin Taylor, Christoph Studer, and Tom Goldstein.
\newblock Visualizing the loss landscape of neural nets.
\newblock In {\em Advances in Neural Information Processing Systems}, pages
  6389--6399, 2018.

\bibitem{ramachandran2017searching}
Prajit Ramachandran, Barret Zoph, and Quoc~V Le.
\newblock Searching for activation functions.
\newblock {\em arXiv preprint arXiv:1710.05941}, 2017.

\bibitem{selvaraju2017grad}
Ramprasaath~R Selvaraju, Michael Cogswell, Abhishek Das, Ramakrishna Vedantam,
  Devi Parikh, Dhruv Batra, et~al.
\newblock Grad-cam: Visual explanations from deep networks via gradient-based
  localization.
\newblock In {\em ICCV}, pages 618--626, 2017.

\bibitem{henriques2015abeta}
Ana~Gabriela Henriques, Joana~Machado Oliveira, Liliana~Patr{\'\i}cia Carvalho,
  and Odete AB da~Cruz e~Silva.
\newblock A$\beta$ influences cytoskeletal signaling cascades with consequences
  to alzheimer’s disease.
\newblock {\em Molecular neurobiology}, 52(3):1391--1407, 2015.

\bibitem{imgaug}
Alexander~B. Jung.
\newblock {imgaug}.
\newblock \url{https://github.com/aleju/imgaug}, 2018.
\newblock [Online; accessed 25-Aug-2019].

\bibitem{berning2015segem}
Manuel Berning, Kevin~M Boergens, and Moritz Helmstaedter.
\newblock Segem: efficient image analysis for high-resolution connectomics.
\newblock {\em Neuron}, 87(6):1193--1206, 2015.

\bibitem{plaza2018analyzing}
Stephen~M Plaza and Jan Funke.
\newblock Analyzing image segmentation for connectomics.
\newblock {\em Frontiers in Neural Circuits}, 12, 2018.

\bibitem{matejeksynapse}
Brian Matejek, Donglai Wei, Xueying Wang, Jinglin Zhao, K{\'a}lm{\'a}n
  Pal{\'a}gyi, and Hanspeter Pfister.
\newblock Synapse-aware skeleton generation for neural circuits.

\bibitem{lyons2014rough}
Terry Lyons.
\newblock Rough paths, signatures and the modelling of functions on streams.
\newblock {\em arXiv preprint arXiv:1405.4537}, 2014.

\bibitem{chevyrev2016primer}
Ilya Chevyrev and Andrey Kormilitzin.
\newblock A primer on the signature method in machine learning.
\newblock {\em arXiv preprint arXiv:1603.03788}, 2016.

\bibitem{graham2013sparse}
Benjamin Graham.
\newblock Sparse arrays of signatures for online character recognition.
\newblock {\em arXiv preprint arXiv:1308.0371}, 2013.

\bibitem{xie2017learning}
Zecheng Xie, Zenghui Sun, Lianwen Jin, Hao Ni, and Terry Lyons.
\newblock Learning spatial-semantic context with fully convolutional recurrent
  network for online handwritten chinese text recognition.
\newblock {\em IEEE transactions on pattern analysis and machine intelligence},
  40(8):1903--1917, 2017.

\bibitem{yang2017leveraging}
Weixin Yang, Terry Lyons, Hao Ni, Cordelia Schmid, Lianwen Jin, and Jiawei
  Chang.
\newblock Leveraging the path signature for skeleton-based human action
  recognition.
\newblock {\em arXiv preprint arXiv:1707.03993}, 2017.

\bibitem{kormilitzin2016application}
AB~Kormilitzin, KEA Saunders, PJ~Harrison, JR~Geddes, and TJ~Lyons.
\newblock Application of the signature method to pattern recognition in the
  cequel clinical trial.
\newblock {\em arXiv preprint arXiv:1606.02074}, 2016.

\bibitem{kormilitzin2017detecting}
Andrey Kormilitzin, Kate~EA Saunders, Paul~J Harrison, John~R Geddes, and Terry
  Lyons.
\newblock Detecting early signs of depressive and manic episodes in patients
  with bipolar disorder using the signature-based model.
\newblock {\em arXiv preprint arXiv:1708.01206}, 2017.

\bibitem{arribas2017signature}
Imanol~Perez Arribas, Kate Saunders, Guy Goodwin, and Terry Lyons.
\newblock A signature-based machine learning model for bipolar disorder and
  borderline personality disorder.
\newblock {\em arXiv preprint arXiv:1707.07124}, 2017.

\bibitem{moore2018using}
PJ~Moore, J~Gallacher, and TJ~Lyons.
\newblock Using path signatures to predict a diagnosis of alzheimer's disease.
\newblock {\em arXiv preprint arXiv:1808.05865}, 2018.

\bibitem{moore2019random}
PJ~Moore, TJ~Lyons, John Gallacher, Alzheimer’s Disease~Neuroimaging
  Initiative, et~al.
\newblock Random forest prediction of alzheimer’s disease using pairwise
  selection from time series data.
\newblock {\em PloS one}, 14(2):e0211558, 2019.

\bibitem{arribas2018derivatives}
Imanol~Perez Arribas.
\newblock Derivatives pricing using signature payoffs.
\newblock {\em arXiv preprint arXiv:1809.09466}, 2018.

\bibitem{gyurko2013extracting}
Lajos~Gergely Gyurk{\'o}, Terry Lyons, Mark Kontkowski, and Jonathan Field.
\newblock Extracting information from the signature of a financial data stream.
\newblock {\em arXiv preprint arXiv:1307.7244}, 2013.

\bibitem{levin2013learning}
Daniel Levin, Terry Lyons, and Hao Ni.
\newblock Learning from the past, predicting the statistics for the future,
  learning an evolving system.
\newblock {\em arXiv preprint arXiv:1309.0260}, 2013.

\bibitem{bonnier2019deep}
Patric Bonnier, Patrick Kidger, Imanol Perez~Arribas, Cristopher Salvi, and
  Terry Lyons.
\newblock Deep signatures.
\newblock {\em arXiv preprint arXiv:1905.08494}, 2019.

\end{thebibliography}

\appendix

\section{List of compounds used for training and screening}
\label{appendix:c}

\begin{table}[ht]
    \centering
    \begin{tabular}{lllll}
        Amprolium &	Bucladesine	& Camptothecin	& Carbimazol & Cefapirin \\
        Chloropyrazine & Chloroquine & Chrysenequinone & Cyclopentolate & Danazol \\
        Diflunisal & Dihydroergotamine & Doxylamine & Ellipticine & Ethisterone \\
        Etidronic Acid & Fulvestrant & Harpagoside & Irinotecan & Isocarboxazid \\
        Levobunolol	& Menadione	& Mephenytoin &	Mercaptopurine & Metaclopramide \\
        Ofloxacin & Orciprenaline &	Oxolamine &	Oxybenzone & Piperlongumine \\
        Puromycin &	Raubasine &	Rifabutin &	Sanguinarine &	Terazosin \\
        Thiamphenicol				 
    \end{tabular}
    \caption{A list of candidate compounds used for training and screening.}
    \label{tab:list_of_compounds}
\end{table}

\section{Grad-CAM visualisation}
\label{appendix:a}

\begin{figure}
    \centering
      \begin{subfigure}[t]{0.48\textwidth}
        \centering
        \includegraphics[width=0.95\linewidth]{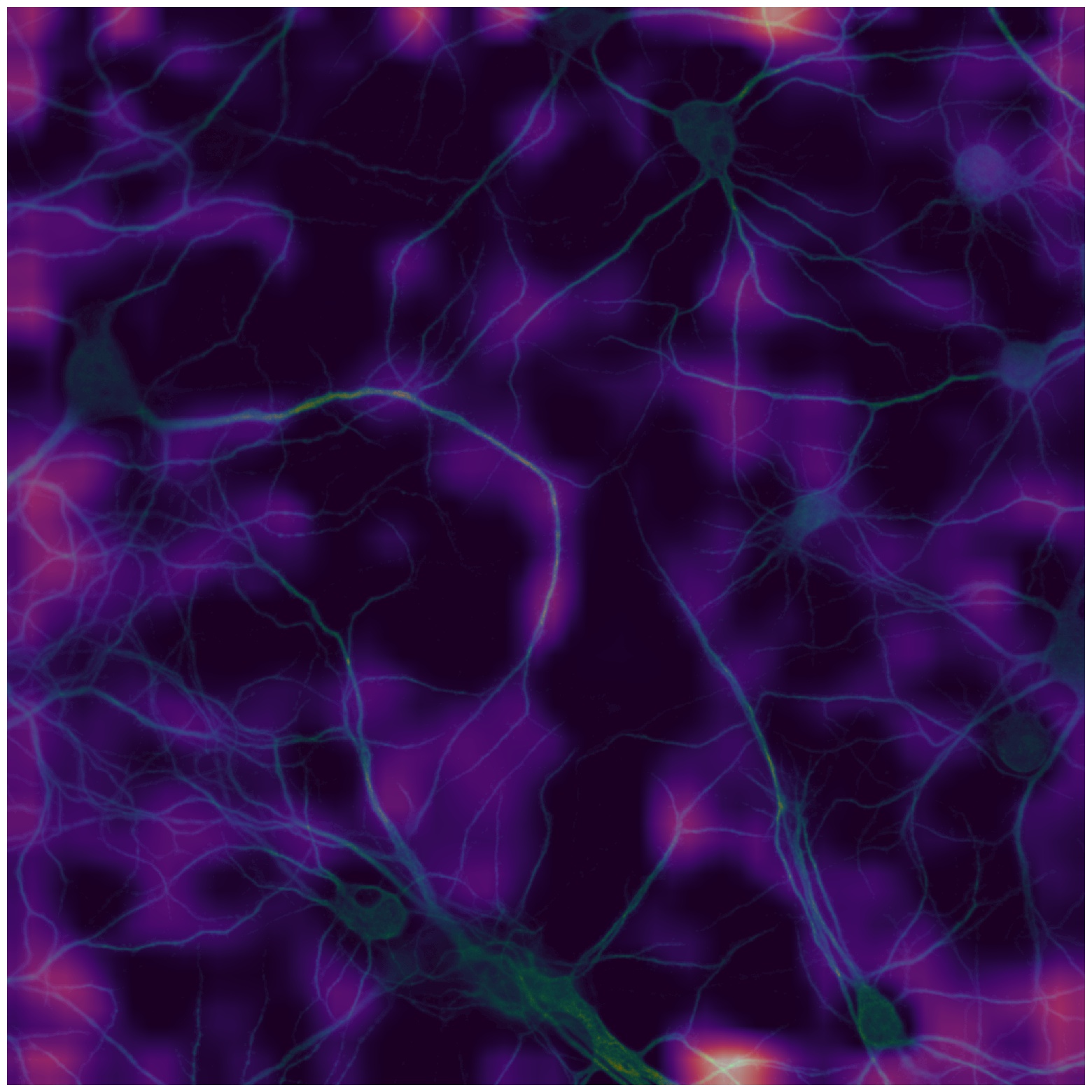}
        \caption{(1, 0, 0.991)}
        \end{subfigure}%
      \begin{subfigure}[t]{0.48\textwidth}
        \centering
        \includegraphics[width=0.95\linewidth]{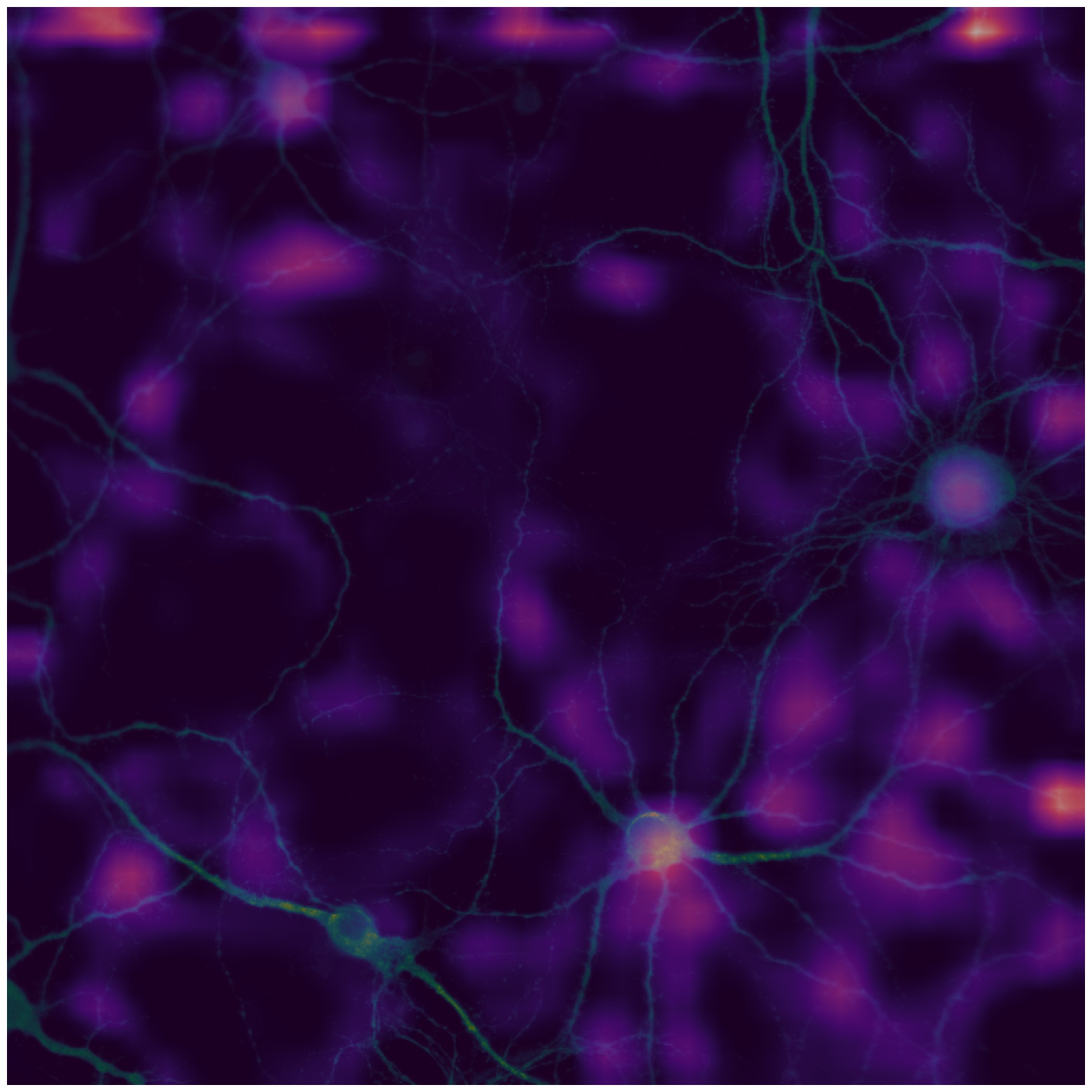}
        \caption{(1, 30, 0.998)}
        \end{subfigure}
    \medskip
    \centering
      \begin{subfigure}[t]{0.48\textwidth}
        \centering
        \includegraphics[width=0.95\linewidth]{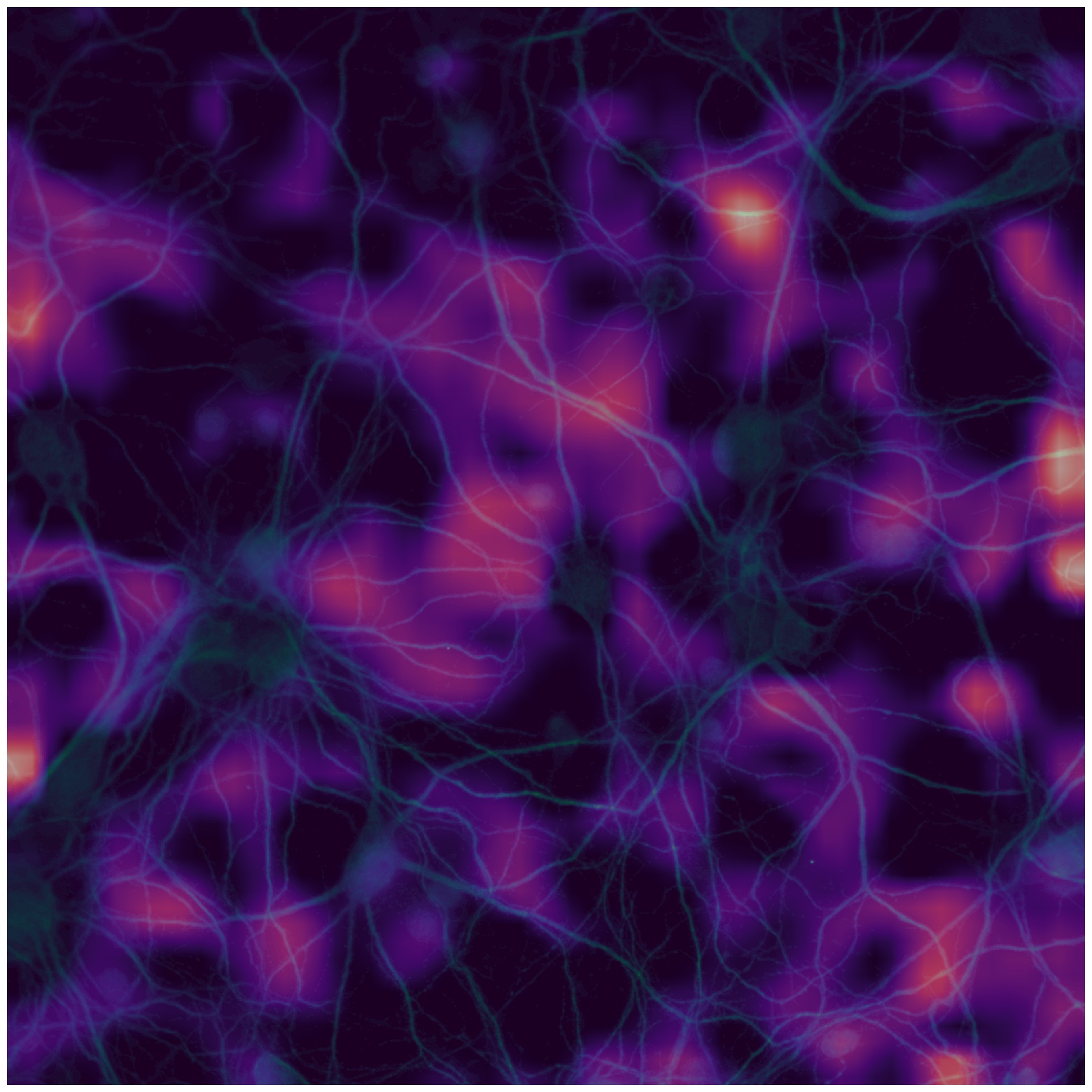}
        \caption{(3, 0, 0.996)}
        \end{subfigure}%
      \begin{subfigure}[t]{0.48\textwidth}
        \centering
        \includegraphics[width=0.95\linewidth]{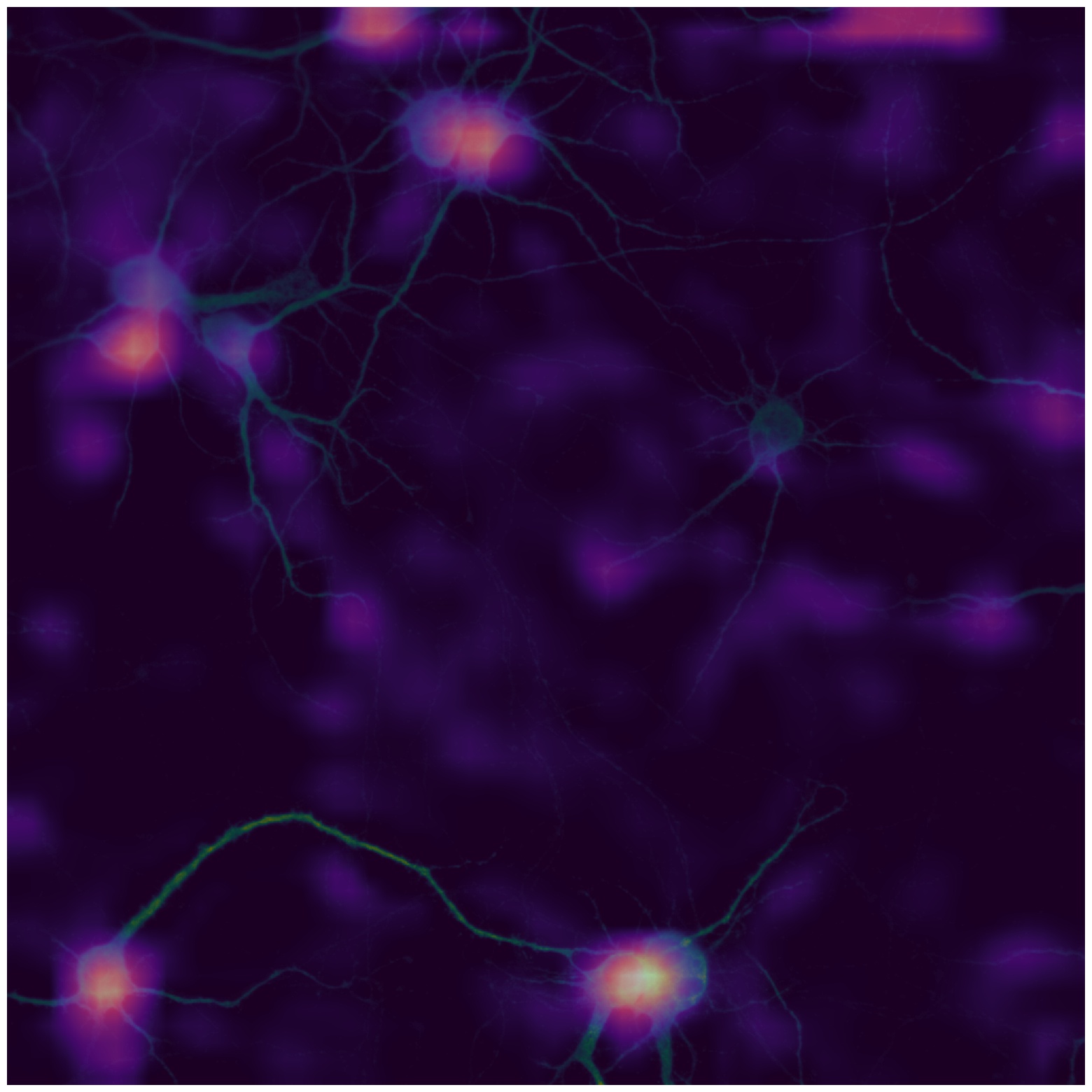}
        \caption{(3, 30, 0.999)}
        \end{subfigure}
    \medskip
    \centering
      \begin{subfigure}[t]{0.48\textwidth}
        \centering
        \includegraphics[width=0.95\linewidth]{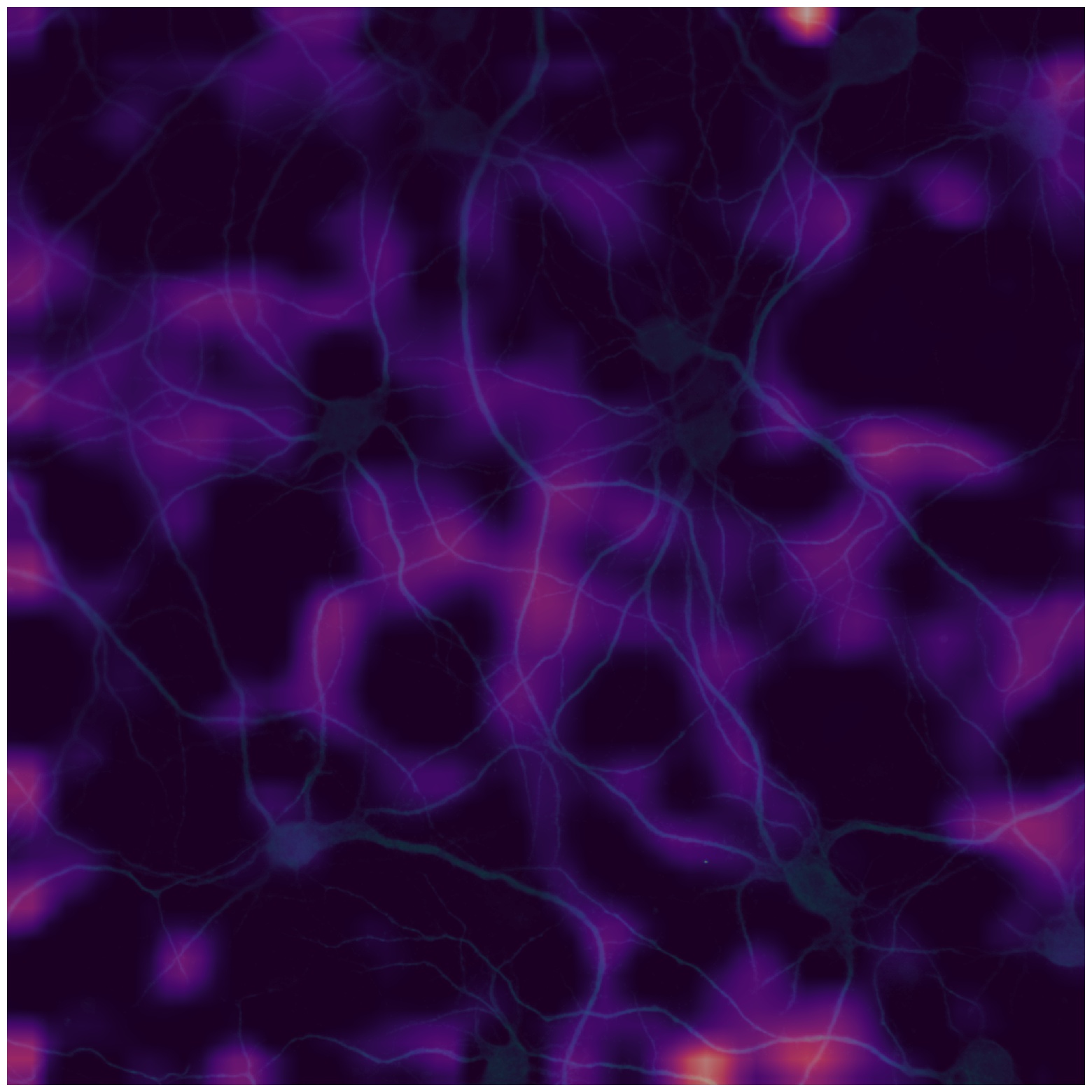}
        \caption{(10, 0, 0.995)}
        \end{subfigure}%
      \begin{subfigure}[t]{0.48\textwidth}
        \centering
        \includegraphics[width=0.95\linewidth]{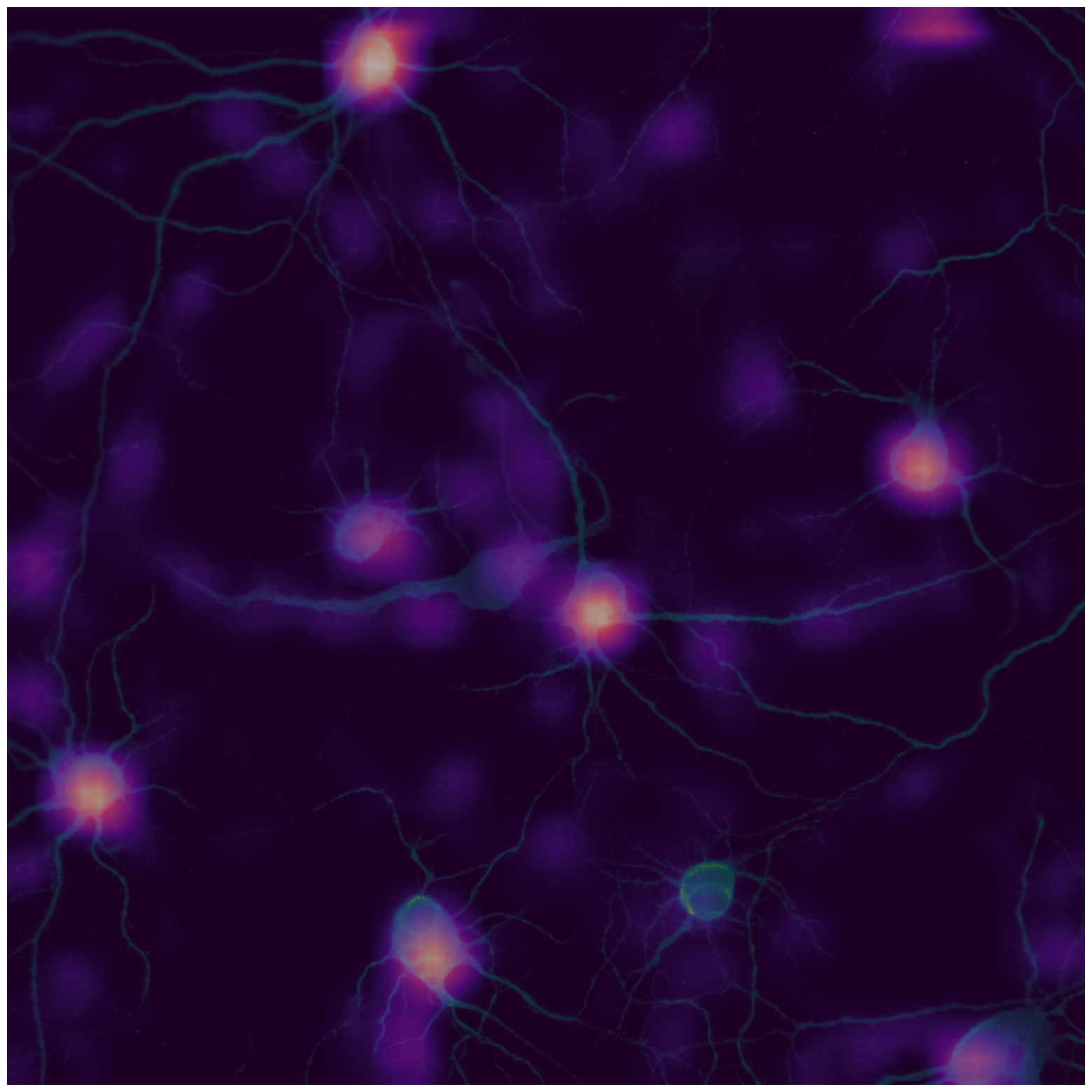}
        \caption{(10, 30, 0.999)}
        \end{subfigure}
    \caption{Random images from the plate with \textbf{Raubasine} compound. Each image is described by a triple (a, b, c), where a - compound dose ($\mu M$), b - $A\beta_{(25-30)}$ dose ($\mu M$) and the predicted score from the model that the image was classified correctly. Left column corresponds to treatments with only Raubasine at various regimes and without added 30 $\mu M$ of $A\beta_{(25-30)}$.}
    \label{fig:raubasine_grad_cam}
\end{figure}

\begin{figure}
    \centering
      \begin{subfigure}[t]{0.48\textwidth}
        \centering
        \includegraphics[width=0.95\linewidth]{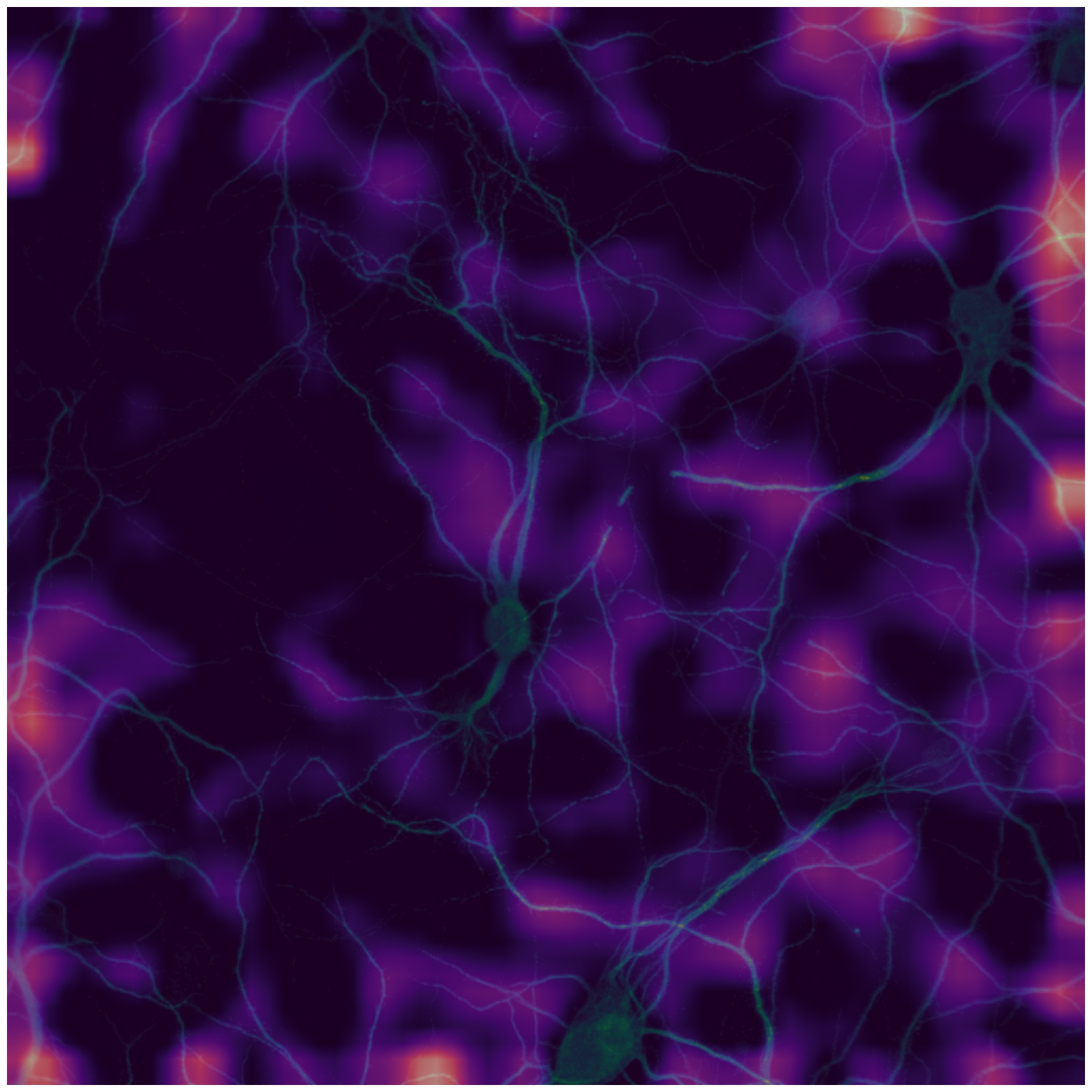}
        \caption{(1, 0, 0.999)}
        \end{subfigure}%
      \begin{subfigure}[t]{0.48\textwidth}
        \centering
        \includegraphics[width=0.95\linewidth]{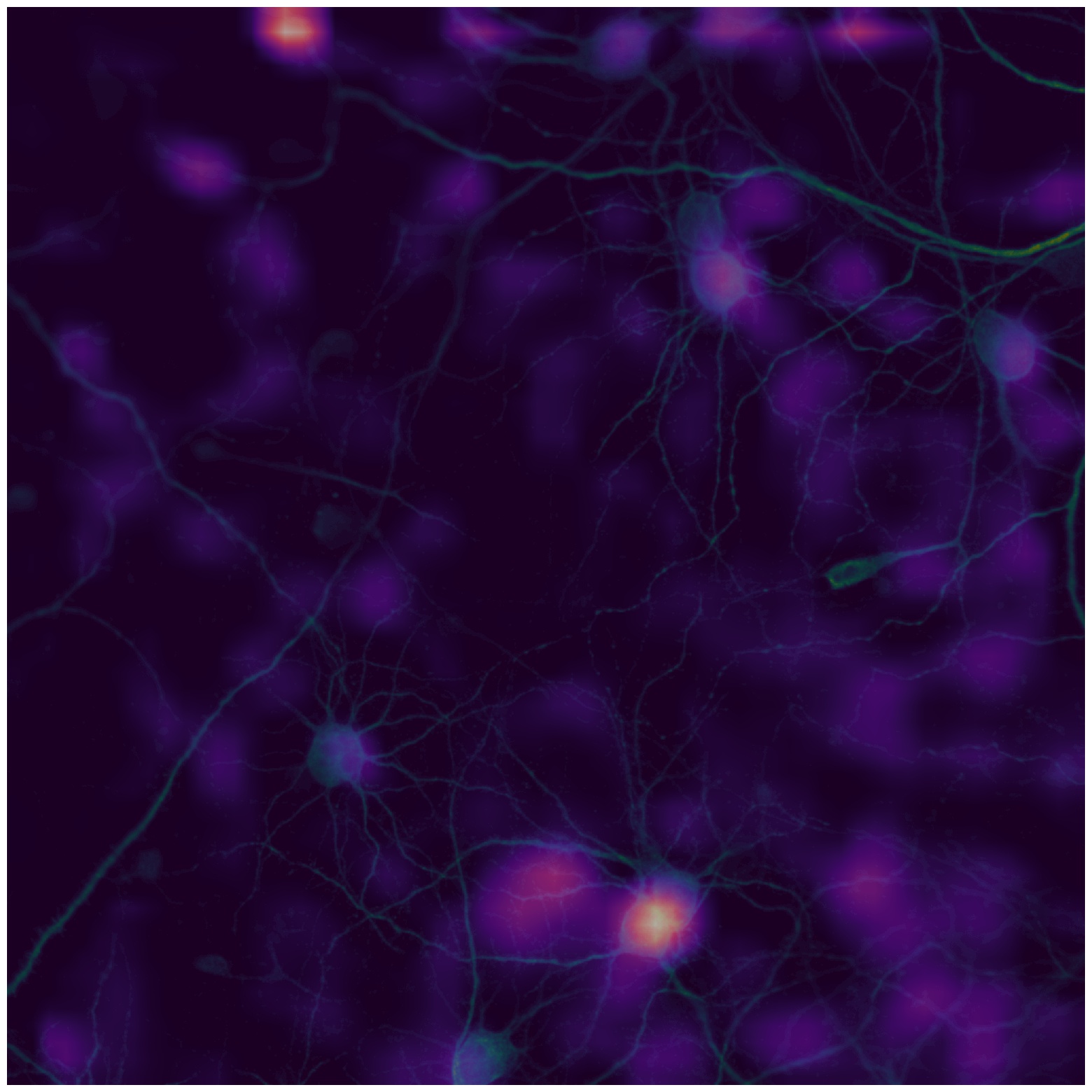}
        \caption{(1, 30, 0.999)}
        \end{subfigure}
    \medskip
    \centering
      \begin{subfigure}[t]{0.48\textwidth}
        \centering
        \includegraphics[width=0.95\linewidth]{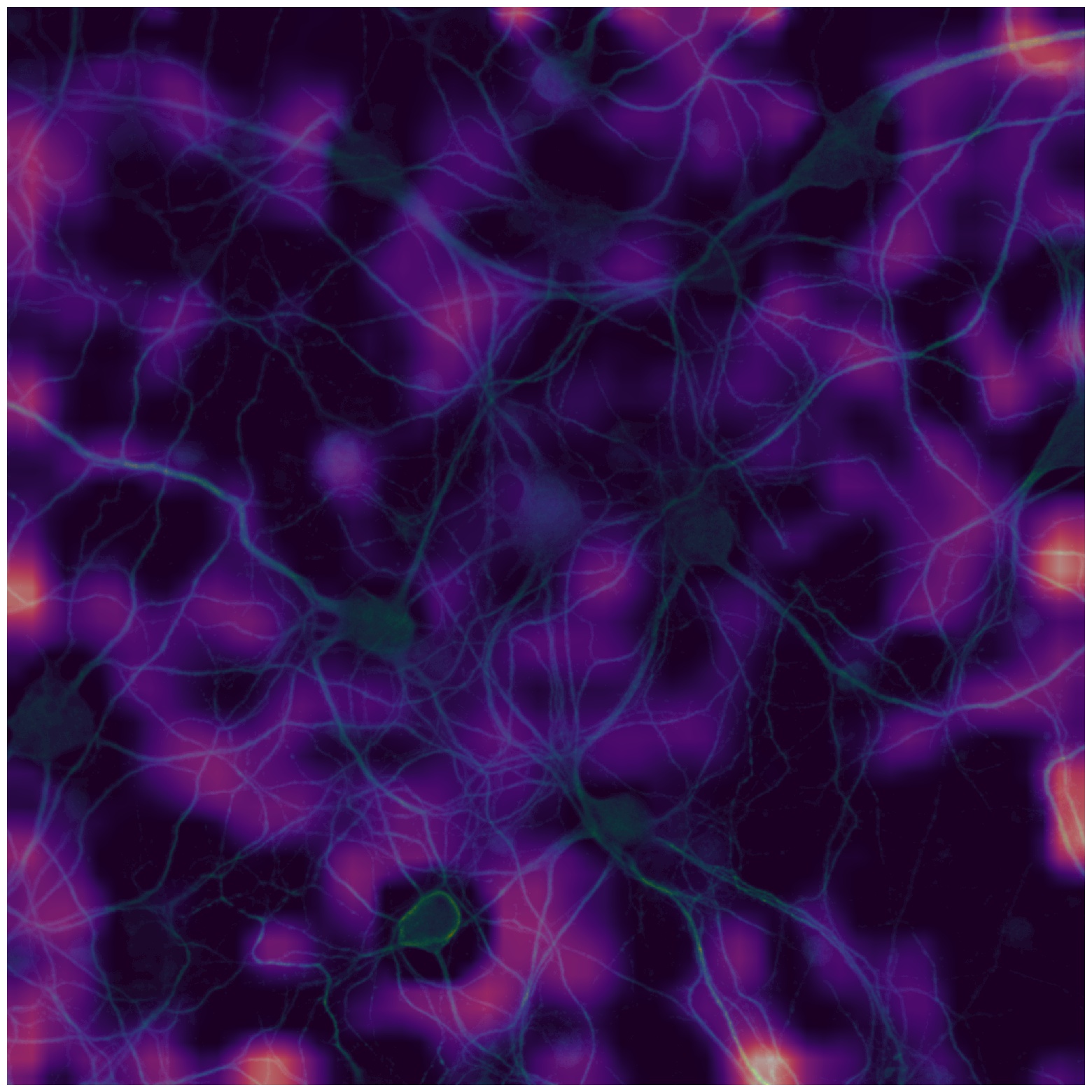}
        \caption{(3, 0, 0.918)}
        \end{subfigure}%
      \begin{subfigure}[t]{0.48\textwidth}
        \centering
        \includegraphics[width=0.95\linewidth]{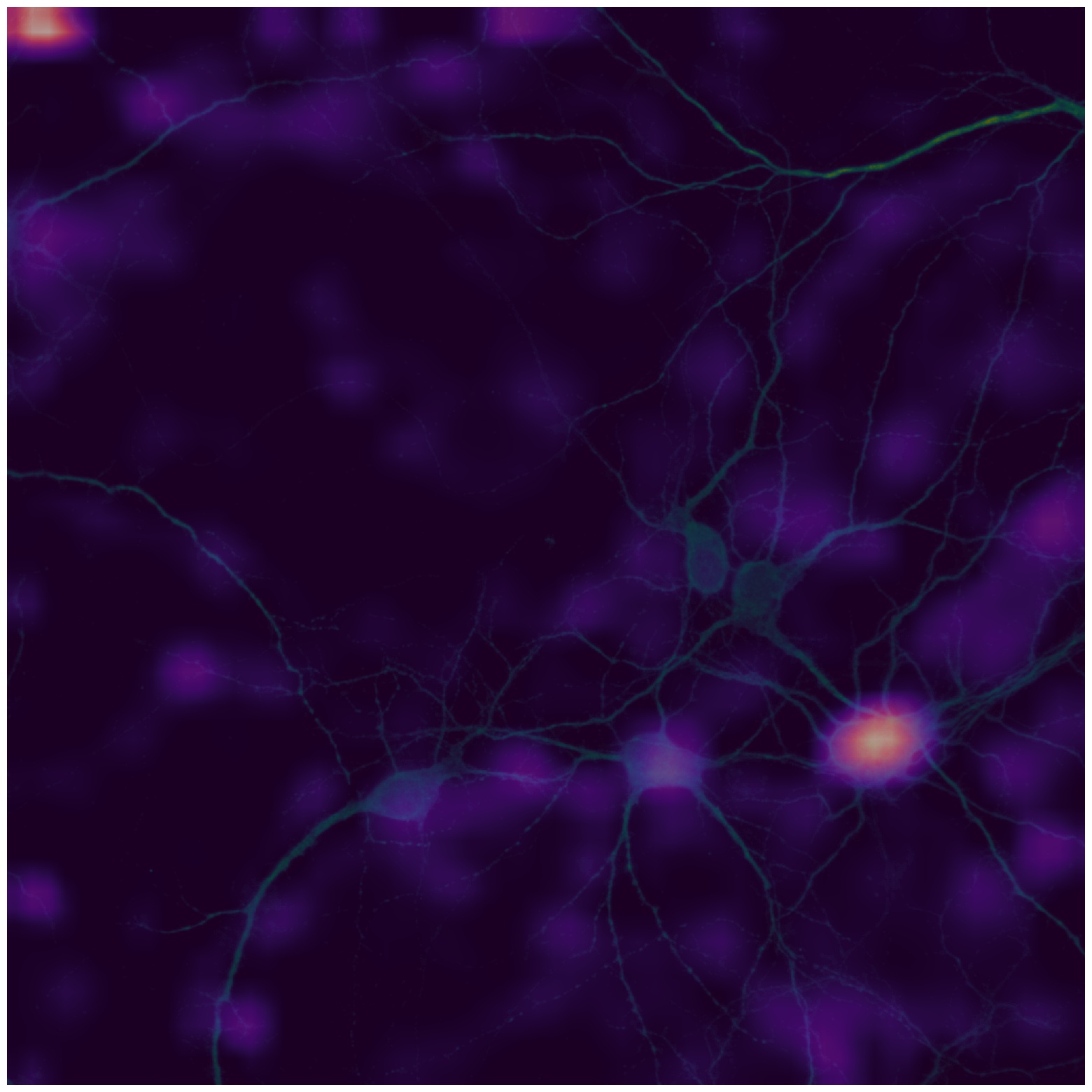}
        \caption{(3, 30, 0.999)}
        \end{subfigure}
    \medskip
    \centering
      \begin{subfigure}[t]{0.48\textwidth}
        \centering
        \includegraphics[width=0.95\linewidth]{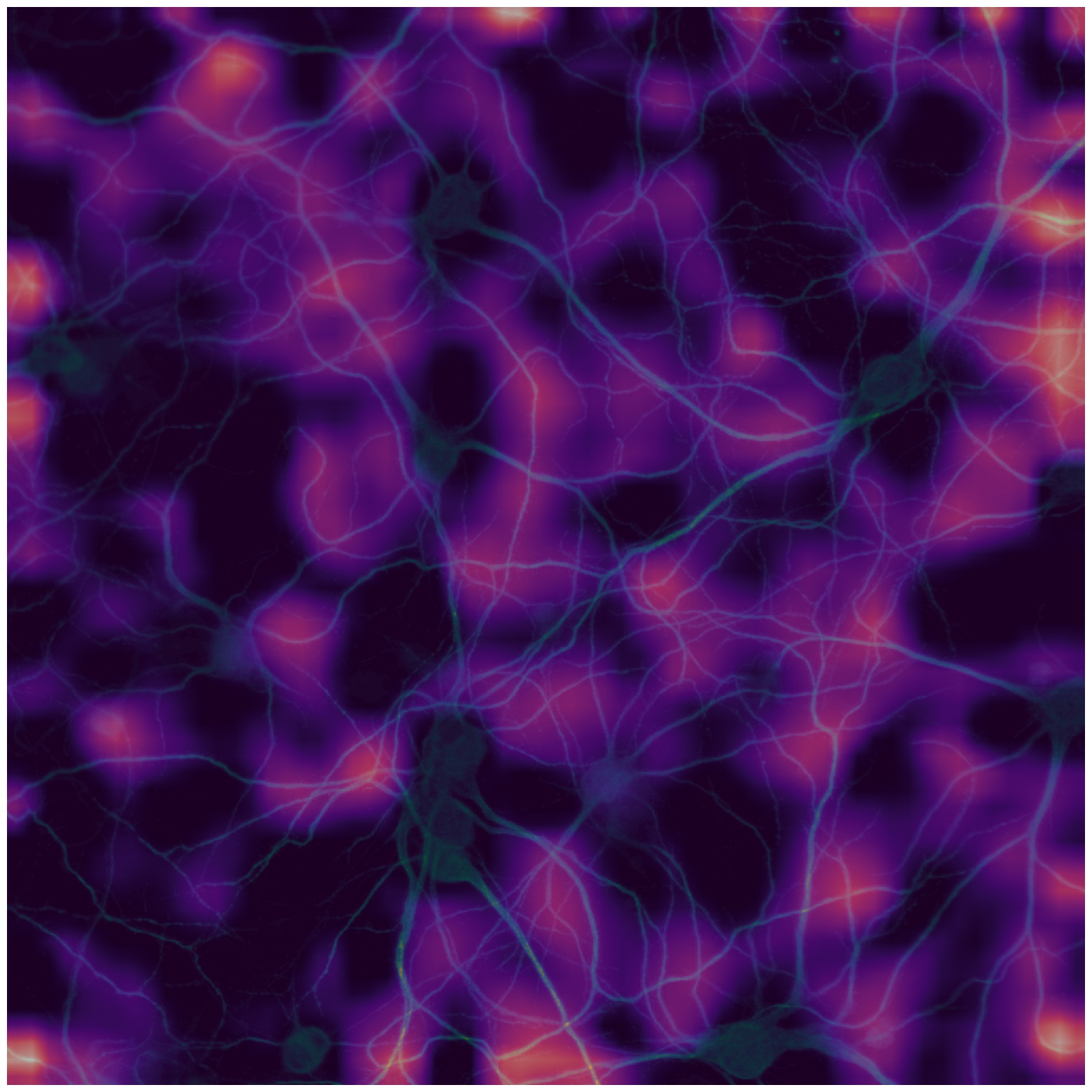}
        \caption{(10, 0, 0.998)}
        \end{subfigure}%
      \begin{subfigure}[t]{0.48\textwidth}
        \centering
        \includegraphics[width=0.95\linewidth]{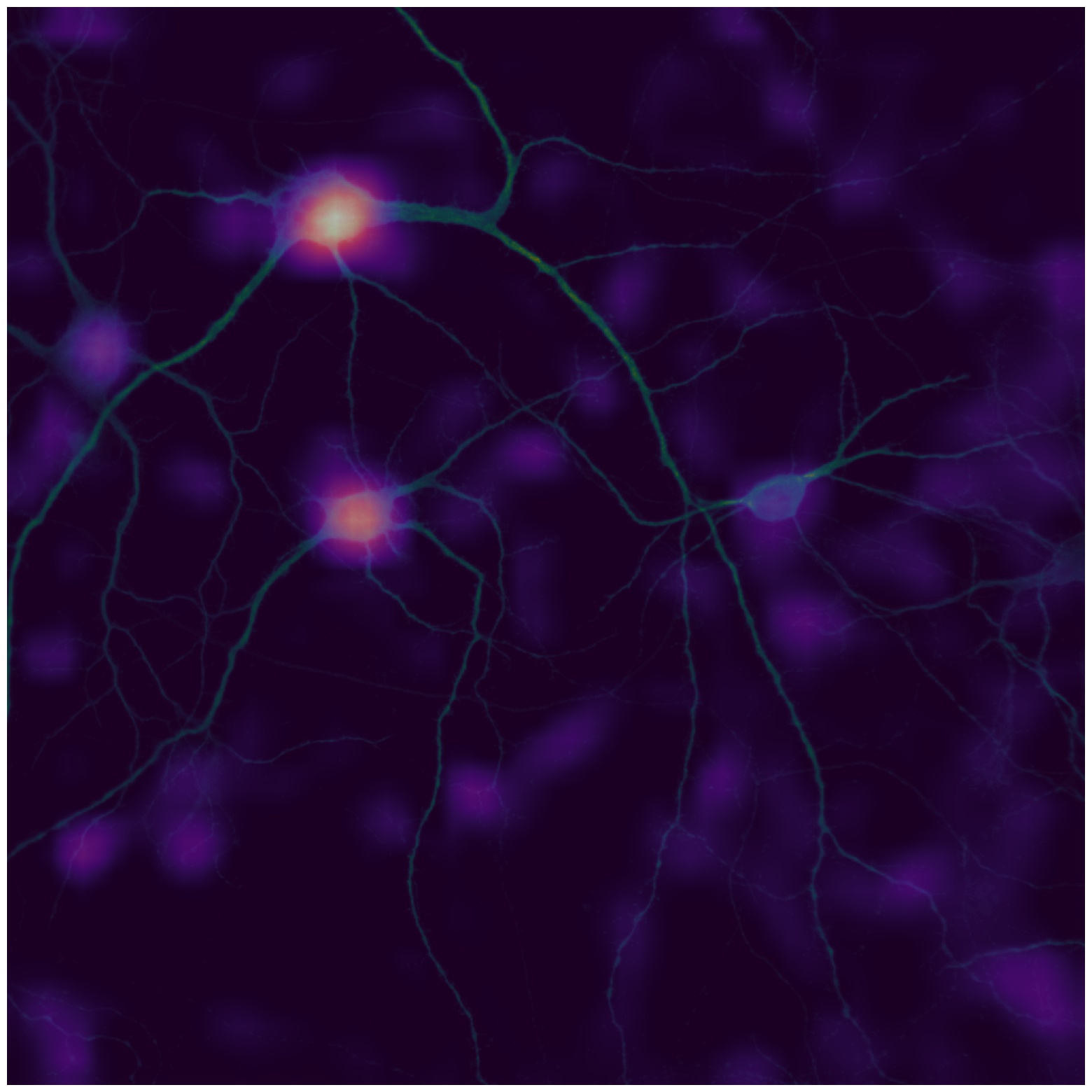}
        \caption{(10, 30, 0.997)}
        \end{subfigure}
    \caption{Random images from the plate with \textbf{Thiamphenicol} compound.  Each image is described by a triple (a, b, c), where a - compound dose ($\mu M$), b - $A\beta_{(25-30)}$ dose ($\mu M$) and the predicted score from the model that the image was classified correctly. Left column corresponds to treatments with only Thiamphenicol at various regimes and without added 30 $\mu M$ of $A\beta_{(25-30)}$.}
    \label{fig:thiamphenicol_grad_cam}
\end{figure}

The model predictions using images of two randomly selected compounds: Raubasine and Thiamphenicol are presented in Fig. \ref{fig:raubasine_grad_cam} and Fig. \ref{fig:thiamphenicol_grad_cam}. The pooled scores of these two compounds over an assay plate are summarised in Tables \ref{tab:raubasine} and \ref{tab:thiamphenicol} respectively.

\newpage
\section{Distribution of predicted scores within wells}\label{appendix:b}

Plots below demonstrate the distribution of predicted scores for each images from within the same well using the two held-out assay plates with Raubasine and Thiamphenicol compounds. Each row corresponds to a single compound treatment and six columns denote replications of the same treatment, following the assay plate structure as shown in Fig. \ref{fig:assay_plate}. 

\end{document}